\documentclass[reprint,aps,pra,preprint,groupedaddress,twocolumn,showpacs,superscriptaddress,amsmath,amssymb]{revtex4-1}

\usepackage{graphicx}
\usepackage{epstopdf}
\usepackage{dcolumn}
\usepackage{multirow}
\usepackage{amsmath}
\usepackage{amsthm}

\usepackage{array}
\usepackage{longtable}
\usepackage{bm,bbm}
\usepackage{amssymb}
\usepackage{subfigure}
\usepackage{epsfig}
\usepackage[T1]{fontenc}
\usepackage{blindtext}
\usepackage[utf8]{inputenc}
\usepackage{float}
\usepackage{rotating}
\usepackage{subfigure}

\makeatletter

\newcommand{\Rmnum}[1]{\expandafter\@slowromancap\romannumeral  #1@}
\makeatother
\begin{document}
\title{Finite-key analysis for a practical decoy-state twin-field quantum key distribution\\}

\author{Shao-Fu He}
\affiliation{Henan Key Laboratory of Quantum Information and Cryptography, SSF IEU, Zhengzhou, Henan 450001, China\\}
\affiliation{Synergetic Innovation Center of Quantum Information and Quantum Physics, University of Science and Technology of China, Hefei, Anhui 230026, China\\}
\author{Yang Wang}\email{wy@qiclab.cn}
\affiliation{Henan Key Laboratory of Quantum Information and Cryptography, SSF IEU, Zhengzhou, Henan 450001, China\\}
\affiliation{Synergetic Innovation Center of Quantum Information and Quantum Physics, University of Science and Technology of China, Hefei, Anhui 230026, China\\}
\author{Hong-Wei Li}
\affiliation{Henan Key Laboratory of Quantum Information and Cryptography, SSF IEU, Zhengzhou, Henan 450001, China\\}
\affiliation{Synergetic Innovation Center of Quantum Information and Quantum Physics, University of Science and Technology of China, Hefei, Anhui 230026, China\\}
\author{Wan-Su Bao}
\email{bws@qiclab.cn}
\affiliation{Henan Key Laboratory of Quantum Information and Cryptography, SSF IEU, Zhengzhou, Henan 450001, China\\}
\affiliation{Synergetic Innovation Center of Quantum Information and Quantum Physics, University of Science and Technology of China, Hefei, Anhui 230026, China\\}

\date{\today}

\begin{abstract}
\fontsize{10pt}{10pt}
Twin-field quantum key distribution (TF-QKD), which is immune to all possible detector side channel attacks, enables two remote legitimate users to perform secure communications without quantum repeaters. With the help of a central node, TF-QKD is expected to overcome the linear key-rate constraint using current technologies. However, the security of the former TF-QKD protocols relies on the hypothesis of infinite-key and stable sources. In this paper, we present the finite-key analysis of a practical decoy-state twin-field quantum key distribution with variant statistical fluctuation models. We examine the composable security of the protocol with intensity fluctuations of unstable sources employing Azuma's inequality. Our simulation results indicate that the secret key rate is able to surpass the linear key-rate bound with limited signal pulses and intensity fluctuations. In addition, the effect of intensity fluctuations is extremely significant for small size of total signals.
\fontsize{10pt}{10pt}
\end{abstract}

\maketitle

\section{Introduction\textbackslash\textbackslash}
\fontsize{10pt}{10pt}
Quantum key distribution (QKD)~\cite{1,2} is considered to be the most mature application of quantum information science. Since the establishment of the first protocol, great efforts have been devoted to develop quantum key distribution~\cite{3,4}. However, the transmission loss of optical pulses, which is an intrinsic property of the quantum channels, has significantly limited the communication distance between the legitimate users~\cite{5,6}.

In order to break the limitation of channel transmittance, the quantum repeater scheme was proposed. Unfortunately, quantum repeaters are impractical to be implemented with current quantum communication technologies~\cite{7,8,9}. On the other side, Lo \textit{et al.}~\cite{10} proposed the measurement-device-independent quantum key distribution (MDI-QKD) protocol to prevent all possible detector side channel attacks. Nevertheless, MDI-QKD cannot remove the bottleneck of long-distance quantum communication either.

Recently, Lucamarini \textit{et al.}~\cite{11} proposed the unprecedented twin-field quantum key distribution (TF-QKD) scheme which was conjectured to overcome the rate-distance limit without trusted relays. Based on single-photon interference at the beamsplitter of an untrusted node, the secret key rate of TF-QKD achieved a quadratic improvement over the traditional phase-encoding scheme of MDI-QKD~\cite{12}. Considering the huge benefit that the secret key rate scales with the square-root of the channel transmittance, several variations of this cutting-edge TF-QKD protocol have been proposed to offer a more rigorous security proof~\cite{13,14,15,16,17,18,19,20}. Subsequently, experiments related to these variant protocols were carried out to prove the feasibility of TF-QKD with current technologies~\cite{21,22,23,24}. Nonetheless, there still remains an inevitable gap between theories and the practical imperfections. Among these experimental limitations, finite-size effect is a nonnegligible character in the estimation of the ultimate secret key rate~\cite{25,26}. Several finite-key analyses have been proposed to study the practicalities of some TF-QKD protocols~\cite{27,28}. Another weakness we cannot neglect is the instability of the photon source. Instead of the stable source employed in the original TF-QKD protocol, the real system always emits photon pulses whose intensities cannot be asymptotically replaced by a constant value~\cite{29,30,31,32}.

In this work, we focus on the practical decoy-state~\cite{33,34,35} TF-QKD scheme proposed by Grasselli \textit{et al.}~\cite{36}. The decoy-state TF-QKD protocol has two main advantages in real-life implementation: (1) This TF-QKD protocol is capable enough to beat the PLOB bound with only two decoy states. (2) The protocol is quite robust against phase misalignments due to the phase randomization of decoy pulses. We utilize different statistical fluctuation analysis models~\cite{37,38,39,40} to perform the parameter estimation step of the two decoy states TF-QKD protocol. Based on the universally composable framework~\cite{41,42,43}, we obtain a tight secret key rate bound with statistical fluctuations. In the case without intensity fluctuations, we make a brief comparison of the final secret key rates estimated by variant statistical fluctuation analysis tools. However, when analyzing the effect of intensity fluctuations, we noticed that the correlation between two detection events cannot be ignored. Hense, we present a tight finite-key analysis utilizing Azuma's inequality~\cite{44,45} to prove the composable security against general attacks with the existence of intensity fluctuations. Through numerical simulations, we investigate the secret key rates of different total signal pulses with statistical fluctuations and intensity fluctuations. The simulation results indicate that intensity fluctuations have a nonnegligible impact on the performance of the practical decoy-state TF-QKD protocol.

The article is organized as follows. In Sec. II, we present the practical decoy-state TF-QKD protocol. In Sec. III, we provide the parameter estimations of the finite-key analysis with or without the existence of intensity fluctuations. In addition, the numerical simulations are demonstrated in Sec. IV. The conclusion of our work is presented in Sec. V.
\fontsize{10pt}{10pt}

\section{Practical decoy state TF-QKD protocol\textbackslash\textbackslash}
\fontsize{10pt}{10pt}
The twin-field type quantum key distribution protocol proposed in \cite{17} is able to overcome the secret key rate bound effectively. According to the two decoy states method proposed in \cite{36}, we constructed the practical form of this TF-QKD protocol in finite-key regime:

\textit{(i) State preparation.} The two legitimate users Alice and Bob choose $Z$-basis ($X$-basis) independently with probability ${P_Z}$ (${P_X} = 1 - {P_Z}$).  If $Z$-basis is chosen, Alice (Bob) prepares the signal pulses with her (his) trusted coherent state source. In this mode, Alice (Bob) firstly generates a secret key bit ${b_A}$ (${b_B}$) randomly selected from the set ${b_A},{b_B} \in \left\{ {0,1} \right\}$. Then, Alice (Bob) prepares a coherent state pulse $\left| {\sqrt {{\mu _Z}} {e^{i\pi {b_A}}}} \right\rangle $ ($\left| {\sqrt {{\mu _Z}} {e^{i\pi {b_B}}}} \right\rangle $) depending on the bit value ${b_A}$ (${b_B}$) with the pre-agreed intensity ${\mu _Z}$. On the other side, if $X$-basis is chosen, Alice (Bob) prepares the decoy pulse with a phase-randomized coherent state. The intensity of the decoy pulse is picked from the set ${\mu _A},{\mu _B} \in \left\{ {{\mu _0},{\mu _1}} \right\}$ with probabilities ${P_{{\mu _0}}}$ and ${P_{{\mu _1}}} = 1 - {P_{{\mu _0}}}$. According to the basic choices, Alice and Bob send their optical pulses to the untrusted third-party Charlie through quantum channels.

\textit{(ii) Measurement.}  In this step, Charlie is supposed to perform an interference measurement with her beamsplitter and record the outcomes of the two threshold detectors. Due to the measurement results, Charlie announces the measurement outcomes of the two detectors ($L$ and $R$) through public channels. The detection events when one and only one of two detectors clicks are called successful detection events.

\textit{(iii) Sifting.} After steps \textit{i-ii} have been repeated for $N$ times, both parties acquire adequate successful detection events for key sifting. Alice and Bob publicly announce their basic and intensity choices of their optical pulses through an authenticated classical channel. If Alice and Bob have selected the same basis, they record these events as successful detection events ${s_Z}$  (${s_X}$) according to which basis they have chosen.

\textit{(iv) Parameter estimation.} When both parties choose $Z$-basis, they retain their bit values corresponding to successful detection events as their raw key bits. Note that Bob always flips his key bit ${b_B}$ if $R$ detector clicks. Their raw key strings are denoted as ${Z_A}$ and ${Z_B}$. According to the successful detection events observed in the sifting step, Alice and Bob can estimate the upper bound of the unknown phase error rate $E_{ph}^Z$ of their raw key bits using our parameter estimation methods provided in Sec. III.

\textit{(v) Error correction.} In order to obtain an identical key bit string, Alice and Bob carry out an information reconciliation scheme. They sacrifice $lea{k_{EC}}$ bits to perform the error correction step. After that, Alice consumes ${\log _2}(1/{\varepsilon _{cor}})$ bits of her string ${Z_A}$ to perform a random two-universal hash function and sends the hash to Bob. If the hash of Bob's string ${Z_B}$ is different from that of ${Z_A}$, they abort the protocol.

\textit{(vi) Privacy amplification.}  To ensure that the information leakage is under control, they exploit a random two-universal hash function to their secret key strings to extract a more private key string with length $l$.
\fontsize{10pt}{10pt}

\section{Parameter estimations\textbackslash\textbackslash}

\subsection{Secrecy analysis}
\fontsize{10pt}{10pt}
To provide a tight finite-key analysis, we exploit the universally composable framework as the benchmark of our security analysis~\cite{26,39}. The final key string that Alice (Bob) obtains after error correction and privacy amplification is denoted as ${Z'_A}$ (${Z'_B}$). Following the definition of composable security, a QKD protocol can be regarded as 'secure' if the two criterions ‘correctness’ and ‘secrecy’ are satisfied. No matter what attacking strategy the eavesdropper Eve employs, the protocol is ${\varepsilon _{cor}}$-correct if the probability that the final key ${Z'_A}$ and ${Z'_B}$ are not identical satisfies
\begin{equation}\label{1}
\begin{aligned}
\Pr [{Z'_A} \ne {Z'_B}] \le {\varepsilon _{cor}}.
\end{aligned}
\end{equation}
The final key is ${\varepsilon _{\sec }}$-secret from Eve if
\begin{equation}\label{2}
\frac{1}{2}{\left\| {{\rho _{{Z'_A}E}} - {U_{{Z'_A}}} \otimes {\rho _E}} \right\|_1} \le {\varepsilon _{\sec }},
\end{equation}
where ${\rho _{{Z'_A}E}}$ is the joint quantum state of Alice’s final key ${Z'_A}$ and Eve, ${U_{{Z'_A}}}$ is the mixed state of all possible values of ${Z'_A}$. ${\left\|  \cdot  \right\|_1}$ denotes the trace norm which indicates that the joint quantum state ${\rho _{{Z'_A}E}}$ is ${\varepsilon _{\sec }}$-close to the ideal case described as ${U_{{Z'_A}}} \otimes {\rho _E}$. In this case, the protocol is called ‘$\varepsilon$-secure’ if it is not only ${\varepsilon _{cor}}$-correct but also ${\varepsilon _{\sec }}$-secret, where $\varepsilon  \ge {\varepsilon _{cor}} + {\varepsilon _{\sec }}$.

Here we obtain the secret key rate $R$ by deriving the private key string length $l$ with the decoy-state method provided in~\cite{36,39}.
According to the universally composable framework definition~\cite{41}, we provide the detailed estimation of the secret key length in Appendix A. The practical TF-QKD protocol in the finite-key regime is ${\varepsilon _{\sec }}$-secret if the length $l$ of the final key satisfies:
\begin{equation}\label{3}
l \le {s_Z}[1 - h(E_{ph}^Z)] - lea{k_{EC}} - 2{\log _2}\frac{1}{{2{\varepsilon _{PA}}}} - {\log _2}\frac{2}{{{\varepsilon _{cor}}}},
\end{equation}
where $h(x) =  - xlo{g_2}x - (1 - x)lo{g_2}(1 - x)$ is the binary Shannon entropy function. Note that $lea{k_{EC}} = {s_Z}fh\left( {E_\mu ^Z} \right)$ represents the bits we sacrifice to perform the error correction step, where $f$ is the error correction inefficiency. Subsequently, the secret key rate can be obtained by $R = l/N$.
\fontsize{10pt}{10pt}

\subsection{Finite-key analysis without intensity fluctuations}
\fontsize{10pt}{10pt}
In our practical TF-QKD protocol, the successful detection events ${s_Z}$, ${s_X}$ and the bit error rate $E_\mu ^Z$ can be directly observed in real experiments. Thus the key issue of our finite-key analysis is the parameter estimation of the unknown phase error rate $E_{ph}^Z$. In this subsection, we demonstrate the derivation of the upper bound of $E_{ph}^Z$ with the successful detection events observed in the sifting step. A tighter security bound is obtained by applying the random sampling method.

For simplicity, we assume that the lossy channel is symmetrical for Alice and Bob. The total number of signal pulses when both parties choose $Z$-basis ($X$-basis) is denoted as ${N_Z}$ (${N_X}$). The successful detection events corresponding to the $X$-basis are denoted as $s_{AB}^X$, the subscripts represent the intensities ${\mu _A}$ and ${\mu _B}$ chosen by Alice and Bob in the decoy-state preparation step. We notice that ${s_X} = \sum {s_{AB}^X}$ is the total amount of successful detection events in the $X$-basis. Note that $s_{AB}^X$ can be directly observed in the sifting step after both parties announced their intensity choices of the decoy pulses.

Here, we denote ${S_{nm}}$ as the set of successful detection events that Alice (Bob) sends out $n$ ($m$) photons in the decoy pulse. We denote ${s_{nm}}$ as the total amount of set ${S_{nm}}$. The probability ${P_{AB\left| {nm} \right.}}$ denotes the conditional probability that Alice (Bob) chooses ${\mu _A}$ (${\mu _B}$) as the intensity of the decoy pulse given that the signal contains $n$ ($m$) photons. The successful detection events $s_{AB,nm}^X$ denotes that Alice (Bob) sends out $n$ ($m$) photons with the intensity choice of the decoy pulse being ${\mu _A}$ (${\mu _B}$). The eavesdropper Eve cannot change the values of ${s_{nm}}$ after Charlie made the announcements. For a set of unknown but fixed values of ${s_{nm}}$, we have that
\begin{equation}\label{4}
\begin{aligned}
\begin{split}
&s_{AB,nm}^X = \sum\limits_{i = 1}^{{s_{nm}}} {\tau _{i,nm}^{AB}},\\
&s_{AB}^X = \sum\limits_{n,m} {\sum\limits_{i = 1}^{{s_{nm}}} {\tau _{i,nm}^{AB}} },
\end{split}
\end{aligned}
\end{equation}
where $\tau _{i,nm}^{AB} = 1$ with probability ${P_{AB\left| {nm} \right.}}$ and otherwise 0. The expectation value of $s_{AB,nm}^X$ with respect to $\tau _{i,nm}^{AB}$ variables can be obtained by
\begin{equation}\label{5}
\begin{aligned}
s_{AB,nm}^{ * ,X} = {P_{AB\left| {nm} \right.}}{s_{nm}}.
\end{aligned}
\end{equation}
Subsequently, the expectation value of $s_{AB}^X$ is given by
\begin{equation}\label{6}
\begin{aligned}
s_{AB}^{ * ,X} = \sum\limits_{n,m} {s_{AB,nm}^{ * ,X}}  = \sum\limits_{n,m} {{P_{AB\left| {nm} \right.}}{s_{nm}}}.
\end{aligned}
\end{equation}
In this case, Eve cannot change the mean value of $s_{AB}^X$ after Charlie's announcements. Thus, the yields and phase error rate estimated by the expectation values of $s_{AB}^X$ cannot be changed either. Let ${Y_{nm}}$ denote the yield when Alice (Bob) sends out $n$ ($m$) photons. We notice that ${s_{nm}} = {N_X}{P_n}{P_m}{Y_{nm}}$, where ${P_n}$ (${P_m}$) is the probability that Alice (Bob) sends out $n$ ($m$) photons in $X$-basis. The expectation value $s_{AB}^{ * ,X}$ can be rewritten as
\begin{equation}\label{7}
\begin{aligned}
s_{AB}^{ * ,X} = {P_A}{P_B}\sum\limits_{n,m} {{P_{n\left| A \right.}}{P_{m\left| B \right.}}{N_X}{Y_{nm}}},
\end{aligned}
\end{equation}
where ${P_{n\left| A \right.}} = {e^{ - {\mu _A}}}{\mu _A}^n/n!$ and ${P_{m\left| B \right.}} = {e^{ - {\mu _B}}}{\mu _B}^m/m!$ are the Poisson distributions of ${\mu _A}$ and ${\mu _B}$.

With the observed values $s_{AB}^X$ obtained in the sifting step, we can calculate the upper bounds and lower bounds of the expectation values $s_{AB}^{ * ,X}$ with our parameter estimation method. Then Eq. (\ref{7}) can be substituted by the following inequality
\begin{equation}\label{8}
\begin{aligned}
\underline {s_{AB}^{ * ,X}}  \le {P_A}{P_B}\sum\limits_{n,m} {{P_{n\left| A \right.}}{P_{m\left| B \right.}}{N_X}{Y_{nm}}}  \le \overline {s_{AB}^{ * ,X}} ,
\end{aligned}
\end{equation}
where $\underline {s_{AB}^{ * ,X}}$ ($\overline {s_{AB}^{ * ,X}}$) is the lower bound (upper bound) of $s_{AB}^{ * ,X}$.

To find the upper bound of the phase error rate $E_{ph}^Z$, one needs to deal the following issues: \textit{(1)} The estimation of $\underline {s_{AB}^{ * ,X}}$ and $\overline {s_{AB}^{ * ,X}}$ with the successful detection events $s_{AB}^X$ observed in the sifting step. \textit{(2)} The estimation of $E_{ph}^Z$ with the decoy-state method and random sampling without replacement method.

The first issue can be solved by applying variant statistical fluctuation models to the observed values $s_{AB}^X$. We exploit Hoeffding's inequality~\cite{37}, the multiplicative Chernoff bound~\cite{39} and the improved Chernoff bound~\cite{40} to obtain a tight key-rate bound in our protocol. Here, we provide the estimation method with the improved Chernoff bound in the main text. The detailed form of Hoeffding's inequality and the multiplicative Chernoff bound can be referred to in Appendix B.

The improved version of the Chernoff bound is expressed as follows. Every single detection event can be regarded as a random variable in our protocol. Let ${\tau _1},{\tau _2}, \ldots ,{\tau _n}$ be a set of $n$ independent Bernoulli random variables where ${\tau _i} \in \left\{ {0,1} \right\}$ and $\Pr \left( {{\tau _i} = 1} \right) = {P_i}$. Let $\tau  = \sum\nolimits_{i = 1}^n {{\tau _i}}$ denote an observed outcome for a given trial. The mean value of the set is denoted by ${\tau ^ * } = \sum\nolimits_{i = 1}^n {{P_i}}$. For a fixed mean value ${\tau ^ * }$, the upper bound and lower bound of the observed value ${\tau}$ satisfy:
\begin{equation}\label{9}
\begin{aligned}
\begin{split}
&\Pr \left[ {\tau  > \left( {1 + \underline \delta  } \right){\tau ^ * }} \right] < {\left[ {\frac{{{e^{\underline \delta  }}}}{{{{\left( {1 + \underline \delta  } \right)}^{\left( {1 + \underline \delta  } \right)}}}}} \right]^{{\tau ^ * }}} = g\left( {\underline \delta  ,{\tau ^ * }} \right),\\
&\Pr \left[ {\tau  < \left( {1 - \overline \delta  } \right){\tau ^ * }} \right] < {\left[ {\frac{{{e^{ - \overline \delta  }}}}{{{{\left( {1 - \overline \delta  } \right)}^{\left( {1 - \overline \delta  } \right)}}}}} \right]^{{\tau ^ * }}} = g\left( { - \overline \delta  ,{\tau ^ * }} \right),
\end{split}
\end{aligned}
\end{equation}
where $\underline \delta   > 0$, $\overline \delta   \in \left( {0,1} \right)$ and $g\left( {x,y} \right) = {\left[ {{e^x}{{\left( {1 + x} \right)}^{ - \left( {1 + x} \right)}}} \right]^y}$. According to the above equations, we can estimate the upper bound and lower bound of the observed value with a fixed expectation value. However, the fixed expectation value ${\tau ^ * }$ mentioned in our finite-key analysis is unknown. We notice that the expectation value ${\tau ^ * }$ can be bounded by  
\begin{equation}\label{10}
\begin{aligned}
\begin{split}
&\Pr \left[ {{\tau ^ * } > {\tau  \mathord{\left/
 {\vphantom {\tau  {\left( {1 - \overline \delta  } \right)}}} \right.
 \kern-\nulldelimiterspace} {\left( {1 - \overline \delta  } \right)}}} \right] < {\left[ {\frac{{{e^{ - \overline \delta  }}}}{{{{\left( {1 - \overline \delta  } \right)}^{1 - \overline \delta  }}}}} \right]^{\overline {{\tau ^ * }} }} = g\left( { - \overline \delta  ,\overline {{\tau ^ * }} } \right),\\
&\Pr \left[ {{\tau ^ * } < {\tau  \mathord{\left/
 {\vphantom {\tau  {\left( {1 + \underline \delta  } \right)}}} \right.
 \kern-\nulldelimiterspace} {\left( {1 + \underline \delta  } \right)}}} \right] < {\left[ {\frac{{{e^{\underline \delta  }}}}{{{{\left( {1 + \underline \delta  } \right)}^{1 + \underline \delta  }}}}} \right]^{\underline {{\tau ^ * }} }} = g\left( {\underline \delta  ,\underline {{\tau ^ * }} } \right),
\end{split}
\end{aligned}
\end{equation}
where $\overline {{\tau ^ * }}  = {\tau  \mathord{\left/
 {\vphantom {\tau  {\left( {1 - \overline \delta  } \right)}}} \right.
 \kern-\nulldelimiterspace} {\left( {1 - \overline \delta  } \right)}}$ and $\underline {{\tau ^ * }}  = {\tau  \mathord{\left/
 {\vphantom {\tau  {\left( {1 + \underline \delta  } \right)}}} \right.
 \kern-\nulldelimiterspace} {\left( {1 + \underline \delta  } \right)}}$ denote the upper bound and lower bound of the expectation value ${\tau ^ * }$. Then the probabilities that ${\tau ^ * }$ exceeds the upper bound $\overline {{\tau ^ * }}$ and the lower bound $\underline {{\tau ^ * }}$ can be denoted as $\overline \varepsilon   = g\left( { - \overline \delta  ,{\tau  \mathord{\left/
 {\vphantom {\tau  {\left( {1 - \overline \delta  } \right)}}} \right.
 \kern-\nulldelimiterspace} {\left( {1 - \overline \delta  } \right)}}} \right)$ and $\underline \varepsilon   = g\left( {\underline \delta  ,{\tau  \mathord{\left/
 {\vphantom {\tau  {\left( {1 + \underline \delta  } \right)}}} \right.
 \kern-\nulldelimiterspace} {\left( {1 + \underline \delta  } \right)}}} \right)$. In this way, given an observed value ${\tau}$ and failure probabilities $\overline \varepsilon$ and $\underline \varepsilon$,  $\overline \delta$ and $\underline \delta$ can be obtained by solving the following equations
 \begin{equation}\label{11}
\begin{aligned}
\begin{split}
&{\left[ {\frac{{{e^{ - \overline \delta  }}}}{{{{\left( {1 - \overline \delta  } \right)}^{1 - \overline \delta  }}}}} \right]^{{\tau  \mathord{\left/
 {\vphantom {\tau  {\left( {1 - \overline \delta  } \right)}}} \right.
 \kern-\nulldelimiterspace} {\left( {1 - \overline \delta  } \right)}}}} = \overline \varepsilon,\\
&{\left[ {\frac{{{e^{\underline \delta  }}}}{{{{\left( {1 + \underline \delta  } \right)}^{1 + \underline \delta  }}}}} \right]^{{\tau  \mathord{\left/
{\vphantom {\tau  {\left( {1 + \underline \delta  } \right)}}} \right.
\kern-\nulldelimiterspace} {\left( {1 + \underline \delta  } \right)}}}} = \underline \varepsilon.
\end{split}
\end{aligned}
\end{equation}

Now we can utilize the improved Chernoff bound to obtain the upper bound $\overline {s_{AB}^{ * ,X}}$ and lower bound $\underline {s_{AB}^{ * ,X}}$ of the successful detection events $s_{AB}^X$:
\begin{equation}\label{12}
\begin{aligned}
\begin{split}
&\overline {s_{AB}^{ * ,X}}  = {{s_{AB}^X} \mathord{\left/
 {\vphantom {{s_{AB}^X} {\left( {1 - \overline \delta  } \right)}}} \right.
 \kern-\nulldelimiterspace} {\left( {1 - \overline \delta  } \right)}},\\
&\underline {s_{AB}^{ * ,X}}  = {{s_{AB}^X} \mathord{\left/
 {\vphantom {{s_{AB}^X} {\left( {1 + \underline \delta  } \right)}}} \right.
 \kern-\nulldelimiterspace} {\left( {1 + \underline \delta  } \right)}}.
\end{split}
\end{aligned}
\end{equation}
where $\overline \delta$ and $\underline \delta$ can be calculated by
\begin{equation}\label{13}
\begin{aligned}
\begin{split}
&{\left[ {\frac{{{e^{ - \overline \delta  }}}}{{{{\left( {1 - \overline \delta  } \right)}^{1 - \overline \delta  }}}}} \right]^{{{s_{AB}^X} \mathord{\left/
 {\vphantom {{s_{AB}^X} {\left( {1 - \overline \delta  } \right)}}} \right.
 \kern-\nulldelimiterspace} {\left( {1 - \overline \delta  } \right)}}}} = \overline \varepsilon,\\
&{\left[ {\frac{{{e^{\underline \delta  }}}}{{{{\left( {1 + \underline \delta  } \right)}^{1 + \underline \delta  }}}}} \right]^{{{s_{AB}^X} \mathord{\left/
 {\vphantom {{s_{AB}^X} {\left( {1 + \underline \delta  } \right)}}} \right.
 \kern-\nulldelimiterspace} {\left( {1 + \underline \delta  } \right)}}}} = \underline \varepsilon.
\end{split}
\end{aligned}
\end{equation}
We notice that Eq. (\ref{13}) is difficult to solve when the observed values $s_{AB}^X$ are large. In this case, we adopt the simplified approximation provided in ~\cite{40} for $s_{AB}^X \ge  - 6\ln \overline \varepsilon$ ($s_{AB}^X \ge  - 6\ln \underline \varepsilon$). The simplified form can be expressed as
\begin{equation}\label{14}
\begin{aligned}
\begin{split}
&\overline \delta   = \frac{{\sqrt {{{\left( {\ln \overline \varepsilon  } \right)}^2} - 8s_{AB}^X\ln \overline \varepsilon  }  - 3\ln \overline \varepsilon  }}{{2\left( {s_{AB}^X + \ln \overline \varepsilon  } \right)}},\\
&\underline \delta   = \frac{{\sqrt {{{\left( {\ln \underline \varepsilon  } \right)}^2} - 8s_{AB}^X\ln \underline \varepsilon  }  - 3\ln \underline \varepsilon  }}{{2\left( {s_{AB}^X + \ln \underline \varepsilon  } \right)}}.
\end{split}
\end{aligned}
\end{equation}

Here, we have obtained the upper bound and lower bound of the expectation values $s_{AB}^{ * ,X}$ with the observables $s_{AB}^X$. In order to address the second issue, the following two steps are involved. On the one hand, we utilize Eq. (\ref{8}) together with the two decoy states method to acquire a tight upper bound of the bit error rate $E_\mu ^X$ in the $X$-basis. Combining the decoy state method with $\overline {s_{AB}^{ * ,X}}$ and $\underline {s_{AB}^{ * ,X}}$, we can obtain the upper bound $\overline {{Y_{n,m}}}$ and lower bound $\underline {{Y_{n,m}}}$ of the yield when Alice and Bob send out $n$ and $m$ photon states seperately. The detailed derivation of the upper and lower bounds of the yields ${Y_{nm}}$ is provided in Appendix C.

According to the estimation method presented in \cite{17}, the upper bound of the bit error rate in the $X$-basis of the TF-QKD protocol can be given by
\begin{widetext}
\begin{equation}\label{15}
\overline {E_\mu ^X}  = \left[ {{{\left( {\sum\limits_{n = 0,m = 0}^\infty  {{K_{2n,2m}}\sqrt {\overline {{Y_{2n,2m}}} } } } \right)}^2} + {{\left( {\sum\limits_{n = 0,m = 0}^\infty  {{K_{2n + 1,2m + 1}}\sqrt {\overline {{Y_{2n + 1,2m + 1}}} } } } \right)}^2}} \right]{{{N_Z}} \mathord{\left/
 {\vphantom {{{N_Z}} {{s_Z}}}} \right.
 \kern-\nulldelimiterspace} {{s_Z}}}.
\end{equation}
Considering that we have employed the two decoy states method in our parameter estimation, we can calculate the upper bound of $E_\mu ^X$ by utilizing the upper bounds of the yields estimated above and asymptotically replace the upper bounds of the other yields by 1. Then we can rewrite Eq. (\ref{15}) as
\begin{equation}\label{16}
\begin{aligned}
\begin{split}
E_\mu ^X \le \overline {E_\mu ^X}  = &[{({K_{0,0}}\sqrt {\overline {{Y_{0,0}}} }  + {K_{0,2}}\sqrt {\overline {{Y_{0,2}}} }  + {K_{2,0}}\sqrt {\overline {{Y_{2,0}}} }  + \sum\limits_{n = 1,m = 1}^\infty  {{K_{2n,2m}}} )^2}\\
&+ {({K_{1,1}}\sqrt {\overline {{Y_{1,1}}} }  + \sum\limits_{n = 0,m = 0}^\infty  {{K_{2n + 1,2m + 1}}}  - {K_{1,1}})^2}]{{{N_Z}} \mathord{\left/
 {\vphantom {{{N_Z}} {{s_Z}}}} \right.
 \kern-\nulldelimiterspace} {{s_Z}}},
\end{split}
\end{aligned}
\end{equation}
where
\begin{equation}\label{17}
\begin{split}
{K_{2n,2m}} &= {{{e^{ - {\mu _z}}}{\mu _z}^{n + m}} \mathord{\left/
 {\vphantom {{{e^{ - {\mu _z}}}{\mu _z}^{n + m}} {\sqrt {(2n)!(2m)!} }}} \right.
 \kern-\nulldelimiterspace} {\sqrt {(2n)!(2m)!} }},\\
{K_{2n + 1,2m + 1}} &= {{{e^{ - {\mu _z}}}{\mu _z}^{n + m + 1}} \mathord{\left/
 {\vphantom {{{e^{ - {\mu _z}}}{\mu _z}^{n + m + 1}} {\sqrt {(2n + 1)!(2m + 1)!} }}} \right.
 \kern-\nulldelimiterspace} {\sqrt {(2n + 1)!(2m + 1)!} }}.
\end{split}
\end{equation}

On the other hand, the phase error rate $E_{ph}^Z$ cannot be simply replaced by the bit error rate $E_\mu ^X$ when the total signal pulses are limited. Considering the influence of statistical fluctuations, we employ the random sampling without replacement method proposed in \cite{46} to provide a tighter bound for the phase error rate $E_{ph}^Z$. Based on an approximate hypergeometric distribution formula, the upper bound of the phase error rate $E_{ph}^Z$ can be described as
\begin{equation}\label{18}
\overline {E_{ph}^Z}  \le E_\mu ^X + \gamma ({s_X},{s_Z},E_\mu ^X,\varepsilon ''),
\end{equation}
with a failure probability $\varepsilon ''$, where
\begin{equation}\label{19}
\begin{aligned}
\gamma ({s_X},{s_Z},E_\mu ^X,\varepsilon '') = \sqrt {\frac{{\left( {{s_X} + {s_Z}} \right)(1 - E_\mu ^X)E_\mu ^X}}{{{s_X}{s_Z}\ln 2}}{{\log }_2}\frac{{{s_X} + {s_Z}}}{{{s_X}{s_Z}{{\varepsilon ''}^2}E_\mu ^X(1 - E_\mu ^X)}}}.
\end{aligned}
\end{equation}
\end{widetext}
\fontsize{10pt}{10pt}

\subsection{Finite-key analysis with intensity fluctuations}
\fontsize{10pt}{10pt}
Apart from the finite-size effect, another imperfection we cannot ignore in our real-life implementation of TF-QKD is the intensity fluctuations of the photon sources. Considering that we have applied the two decoy states method to our TF-QKD system, the intensity fluctuations of both signal and decoy pulses should be taken into account simultaneously. In the case without intensity fluctuations, we assume that the detection events are independent. However, when the photon sources are unstable, the intensity of an optical pulse might correlate with other pulses if the eavesdropper Eve adopts coherent attacks. In this case, the independent condition of the detection events is not satisfied. To estimate the secret key rate with intensity fluctuations of the photon sources, we exploit Azuma's inequality~\cite{44,45} to perform the finite-key analysis with dependent samples.

According to our practical TF-QKD protocol, three different intensities are utilized for the signal and decoy pulses. For simplicity, we suppose that the intensity fluctuation magnitudes of the signal pulses and the decoy pulses are equal and symmetric for Alice and Bob
\begin{equation}\label{20}
\begin{split}
&{\mu _z}(1 - {\delta _\mu }) = \underline {{\mu _z}}  \le {\mu _z} \le \overline {{\mu _z}}  = {\mu _z}(1 + {\delta _\mu }),\\
&{\mu _i}(1 - {\delta _\mu }) = \underline {{\mu _i}}  \le {\mu _i} \le \overline {{\mu _i}}  = {\mu _i}(1 + {\delta _\mu }),
\end{split}
\end{equation}
where $\mu _i$ is the intensity of the decoy pulse with $i \in \left\{ {0,1} \right\}$, and ${\delta _\mu }$ represents the fluctuation magnitude of the intensities.

Here, Azuma’s inequality~\cite{44} is leveraged to measure the upper bound and lower bound of the expectation value of $s_{AB}^X$ for dependent events. Let $n_A^0,n_A^1, \ldots$ denotes a sequence of random variables. Consider the following two conditions:

(1) \textit{Martingale.} The sequence is called a martingale if and only if it satisfies $E\left[ {n_A^{i + 1}|n_A^0,n_A^1, \ldots ,n_A^i} \right] = n_A^i$ for all non-negative integer $i$, where $E\left[ {n_A^{i + 1}|n_A^0,n_A^1, \ldots ,n_A^i} \right]$ is the expectation value of ${n_A^{i + 1}}$ conditioned on the first $i + 1$ outcomes of the sequence ${n_A^0,n_A^1, \ldots ,n_A^i}$.

(2) \textit{Bounded difference condition.} The sequence satisfies the bounded difference condition if there exists ${c^i} > 0$ such that $\left| {n_A^{i + 1} - n_A^i} \right| \le {c^i}$ for all non-negative integer $i$.

For a sequence $n_A^0,n_A^1, \ldots ,n_A^N$ which contains $N+1$ trials and satisfies the above conditions with ${c^i} = 1$, $n_A^N$ can be bounded by Azuma’s inequality
\begin{equation}\label{21}
\begin{aligned}
\Pr \left[ {\left| {n_A^N - n_A^0} \right| > N\delta } \right] \le 2{e^{ - {{N{\delta ^2}} \mathord{\left/
 {\vphantom {{N{\delta ^2}} 2}} \right.
 \kern-\nulldelimiterspace} 2}}}
\end{aligned}
\end{equation}
for $\delta  \in \left( {0,1} \right)$.

Then, let $\eta _A^0,\eta _A^1, \ldots ,\eta _A^N$ be a set of $N+1$ random but dependent variables that $\eta _A^i \in \left\{ {0,1} \right\}$. We define the $i$th trial of the sequence as
\begin{equation}\label{22}
\begin{aligned}
n_A^i = \left\{ \begin{array}{l}
\tau _A^i - \tau _A^{ * ,i}{\rm{ + }}\eta _A^0{\rm{ {\kern 1pt}  {\kern 1pt}  {\kern 1pt}{\kern 1pt}{\kern 1pt}  {\kern 1pt}  {\kern 1pt}{\kern 1pt} if {\kern 1pt}  {\kern 1pt}  {\kern 1pt}{\kern 1pt} }}i \ge 1{\rm{  }}\\
\eta _A^0{\rm{     {\kern 1pt}  {\kern 1pt}  {\kern 1pt}{\kern 1pt}{\kern 1pt}  {\kern 1pt}  {\kern 1pt}{\kern 1pt}{\kern 1pt}  {\kern 1pt}  {\kern 1pt}{\kern 1pt}{\kern 1pt}  {\kern 1pt}  {\kern 1pt}{\kern 1pt}{\kern 1pt}  {\kern 1pt}  {\kern 1pt}{\kern 1pt}{\kern 1pt}  {\kern 1pt}  {\kern 1pt}{\kern 1pt}{\kern 1pt}  {\kern 1pt}  {\kern 1pt}{\kern 1pt}{\kern 1pt}  {\kern 1pt}  {\kern 1pt}{\kern 1pt}{\kern 1pt}  {\kern 1pt}  {\kern 1pt}{\kern 1pt}{\kern 1pt}  {\kern 1pt}  {\kern 1pt}{\kern 1pt}{\kern 1pt}  {\kern 1pt}  {\kern 1pt}{\kern 1pt}{\kern 1pt}  {\kern 1pt}  {\kern 1pt}{\kern 1pt}{\kern 1pt}  {\kern 1pt} {\kern 1pt} {\kern 1pt}  {\kern 1pt}{\kern 1pt}       if {\kern 1pt}  {\kern 1pt}  {\kern 1pt}{\kern 1pt} }}i = 0
\end{array} \right.
\end{aligned}
\end{equation}
where $\tau _A^i = \sum\limits_{j = 1}^i {\eta _A^j}$ is the observed value of the set $\eta _A^1, \ldots ,\eta _A^N$ with $N$ dependent variables, $\tau _A^{ * ,i} = \sum\limits_{j = 1}^i {P\left( {\eta _A^j|\eta _A^0, \ldots ,\eta _A^{j - 1}} \right)}$ is the expectation value of the set $\eta _A^1, \ldots ,\eta _A^N$. We find that the sequence defined by Eq. (\ref{21}) is a Martingale and satisfies the bounded difference condition with ${c^i} = 1$. Hence, Azuma’s inequality can be applied to the sequence. Then, the expectation value $\tau _A^{ * ,N}$ can be bounded by
\begin{equation}\label{23}
\begin{aligned}
\Pr \left[ {\left| {\tau _A^N - \tau _A^{ * ,N}} \right| > N\delta } \right] \le 2{e^{ - {{N{\delta ^2}} \mathord{\left/
 {\vphantom {{N{\delta ^2}} 2}} \right.
 \kern-\nulldelimiterspace} 2}}}.
\end{aligned}
\end{equation}
Utilizing Azuma’s inequality, we can obtain the upper bound and lower bound of the expectaion value $\tau _A^{ * ,N}$
\begin{equation}\label{24}
\begin{aligned}
\begin{split}
&\overline {\tau _A^{ * ,N}}  = \tau _A^N + {{\hat \Delta }_A},\\
&\underline {\tau _A^{ * ,N}}  = \tau _A^N - {\Delta _A}.
\end{split}
\end{aligned}
\end{equation}
with failure probabilities ${{\hat \varepsilon }_A}$ and ${\varepsilon _A}$, where ${{\hat \Delta }_A} = {g_A}\left( {N,{{\hat \varepsilon }_A}} \right)$, ${\Delta _A} = {g_A}\left( {N,{\varepsilon _A}} \right)$ and ${g_A}\left( {x,y} \right) = \sqrt {2x\ln \left( {1/y} \right)}$.
In our practical TF-QKD protocol, a set of $N$ signal pulses can be regarded as the set $\eta _A^1, \ldots ,\eta _A^N$, where $\eta _A^i = 1$ corresponding to a successful detection event. In this way, we can calculate $\overline {s_{AB}^{ * ,X}}$ and $\underline {s_{AB}^{ * ,X}}$ with the observables utilizing Azuma's inequality.

In our parameter estimation, we only consider the worst case through numerically minimizing the key rate over all the possible intensity choices given by the fluctuation magnitude. According to the bounds given above, we can estimate the upper bound of the bit error rate $E_\mu ^X$ in the $X$-basis. With the same method proposed in the previous subsection, we can obtain the upper bound of $E_{ph}^Z$. The length of the final secret key is determined by solving Eq. (\ref{3}).
\fontsize{10pt}{10pt}

\section{Numerical simulations\textbackslash\textbackslash}
\fontsize{10pt}{10pt}

In this section, we simulate the performance of the practical TF-QKD protocol with finite key length and intensity fluctuations. We calculate the secret key rate against the overall loss ($ - 10{\log _{10}}\eta$) which is measured in $dB$. Particularly, the overall loss consists of the transmission loss of the optical channels and the detection efficiencies of the two detectors. For simplicity, the intensities of the decoy states are set to ${\mu _0} = 0.4$ and ${\mu _1} = {10^{ - 5}}$ with probabilities ${P_{{\mu _0}}} = 0.15$ and ${P_{{\mu _1}}} = 1 - {P_{{\mu _0}}}$. We fix ${\varepsilon _{\sec }} = {10^{ - 10}}$ and ${\varepsilon _{cor}} = {10^{ - 12}}$.

In our practical TF-QKD scheme, phase misalignment does not affect the phase error rates due to the phase randomization of the decoy pulses. Thus, the protocol is robust against small phase misalignments. In this case, the polarization and phase misalignments of the optical pulses after traveling through the quantum channels are fixed to $2\%$ which is the same as the original practical TF-QKD protocol~\cite{36}. The successful detection events $s_{AB}^X$, ${s_Z}$ and the bit error rate in $Z$-basis $E_\mu ^Z$ are directly obtained by the legitimate users after sifting step in real experiments. Here, these observed values are simulated by the linear channel loss model provided in Appendix D.

In Fig. 1, we simulate the secret key rate of the practical TF-QKD with finite-key size using the multiplicative Chernoff bound. Here, we demonstrate the performance of the protocol with different amounts of total signals. As shown in Fig. 1, the protocol is capable enough to beat the PLOB bound with a total photon number of ${10^{12}}$ even when the dark count rate of the detectors is ${P_d} = {10^{ - 7}}$. However, the finite-key effect is more significant if the dark count rate is smaller.

\begin{figure}
  \centering
  \includegraphics[width=8cm]{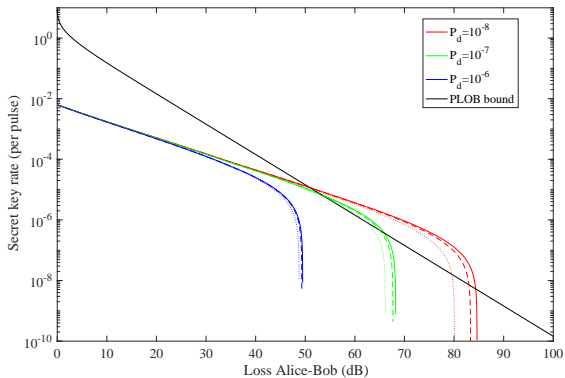}
  \caption{{(Color online) Secret key rates in logarithmic scale against the overall loss for $N = {10^x}$ with $x = 12,13,14$. The solid lines, dashed lines and dotted lines relate to the cases when the dark count rates ${P_d}$ of the detectors are ${10^{ - 8}}$, ${10^{ - 7}}$ and ${10^{ - 6}}$. The solid black line represents the PLOB bound.}}
  \label{1}
\end{figure}

 To figure out the tightest analytical bound of our practical TF-QKD protocol, we compare the performances of Hoeffding's inequality~\cite{37}, the multiplicative Chernoff bound~\cite{39} and the improved Chernoff bound~\cite{40}. As shown in Fig. 2, the results demonstrate that the secret key rate estimated by the improved Chernoff bound is the tightest among three different methods. However, the advantages are diminutive especially when the total number of signal pulses is large.

\begin{figure}
  \centering
  \includegraphics[width=8cm]{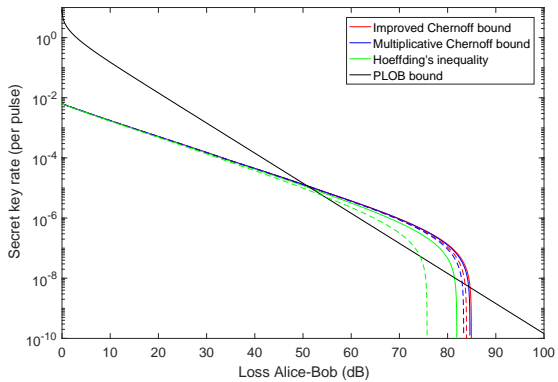}
  \caption{{(Color online) Secret key rates against the overall loss for variant security bounds with different amount of total signal pulses. The green, blue and red lines represents the parameter estimation utilizing Hoeffding’s inequality, the multiplicative Chernoff bound and the improved Chernoff bound. As a comparison, we consider two different total numbers of signal pulses $N = {10^{13}}$ (dashed lines) and $N = {10^{14}}$ (solid lines).}}
  \label{2}
\end{figure}

In the case that the photon sources are instable, we evaluate the secret key rate with different intensity fluctuation magnitudes. Our simulation results indicate that the influence of intensity fluctuations is nonnegligible in real-life implementations. Here, the dark count rate is fixed to ${P_d} = {10^{ - 8}}$. Fig. 3 shows that if ${\delta _\mu }$ is too large, the protocol cannot beat the PLOB bound.

\begin{figure}
  \centering
  \includegraphics[width=8cm]{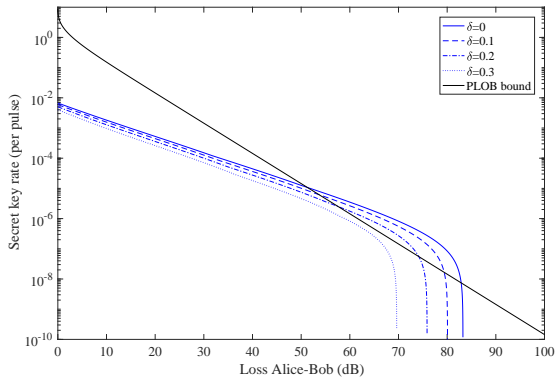}
  \caption{{(Color online) Secret key rates against the overall loss for different fluctuation magnitudes of the intensities with instable sources. The total number of signal pulses is fixed to $N = {10^{15}}$. Accordingly, the curves from left to right are obtained by employing different intensity fluctuation magnitudes (${\delta _\mu } = 0,0.1,0.2,0.3$ separately).}}
  \label{3}
\end{figure}

In order to evaluate the effect of the total amount of signal pulses in the presence of intensity fluctuations, we estimate the secret key rates with different number of total signal pulses given a fixed ${\delta _\mu }$. The results shown in Fig. 4 indicate that the effect of ${\delta _\mu }$ is extremely significant when the data sizes of signal pulses are small.

\begin{figure}
  \centering
  \includegraphics[width=8cm]{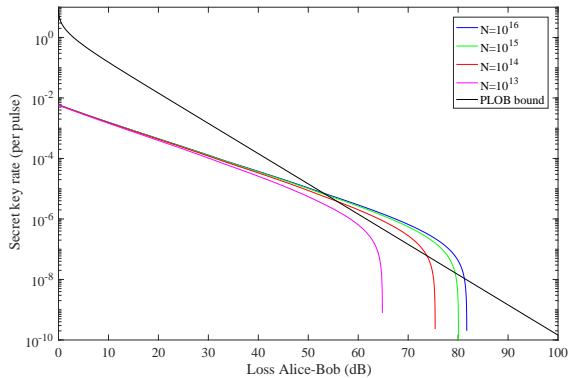}
  \caption{{(Color online) Secret key rates against the overall loss for different data sizes of signal pulses with the intensity fluctuation magnitude ${\delta _\mu }$ fixed to $0.1$. the curves from left to right are obtained for different data sizes ($N = {10^x}$ with $x = 13,14,15,16$ separately).}}
  \label{4}
\end{figure}
\fontsize{10pt}{10pt}

\section{Conclusion\textbackslash\textbackslash}
\fontsize{10pt}{10pt}
In this article, we analyzed the performance of the practical TF-QKD protocol with finite-key effect and intensity fluctuations. Based on the symmetrical assumption of the system, we derived the secret key length formula utilizing the universally composable framework. An estimation of the statistical fluctuations has been provided to characterize the expected values of the successful detection events in the $X$-basis with the observed values obtained in the sifting step. Particularly, comparing variant security bounds, we find that the improved Chernoff bound is the tightest among Hoeffding’s inequality, the multiplicative Chernoff bound and the improved Chernoff bound. In addition, we exploit Azuma's inequality to perform the finite-key analysis with intensity fluctuations. We examined the secret key rates for different total photon pulses and fluctuation magnitudes separately. According to numerical simulation results, we are convinced that the stability of the photon source is essential to the performance of the practical TF-QKD systems especially when the data sizes are relatively smaller. In conclusion, our results of these practical issues might provide an available reference for the real-life implementation of the decoy-state TF-QKD scheme.
\fontsize{10pt}{10pt}

\begin{acknowledgments}
\fontsize{10pt}{10pt}
This work is supported by the National Key Research and Development Program of China (Grant No. 2016YFA0302600) and National Natural Science Foundation of China (Grants No. 61675235, 61605248, 61505261).
\fontsize{10pt}{10pt}
\end{acknowledgments}

\begin{appendix}

\section{Secret key length}
\fontsize{10pt}{10pt}
Here, the whole information Eve learned about Alice's raw key string ${Z_A}$ is summarized by $E'$. By exploiting a random two universal hash function, a ${\varepsilon _{\sec }}$-secret key with length $l$ can be extracted from ${Z_A}$
\begin{equation}\label{A1}
{\varepsilon _{\sec }}= 2\varepsilon ' + \frac{1}{2}\sqrt {{2^{l - H_{\min }^{\varepsilon '}\left( {\left. {{Z_A}} \right|E'} \right)}}},
\end{equation}
where $\varepsilon '$ is the smoothing parameter, $H_{\min }^{\varepsilon '}\left( {\left. {{Z_A}} \right|E'} \right)$ is the smooth min-entropy characterizing the average probability of Eve correctly guessing ${Z_A}$ with the leaked information $E'$~\cite{47}.
 In the error correction step, a maximum amount of $lea{k_{EC}} + {\log _2}(1/{\varepsilon _{cor}})$ bits of the raw key are leaked to Eve. According to the chain rules~\cite{48}, the smooth min-entropy $H_{\min }^{\varepsilon '}\left( {\left. {{Z_A}} \right|E'} \right)$ can be obtained by
\begin{equation}\label{A2}
H_{\min }^{\varepsilon '}\left( {\left. {{Z_A}} \right|E'} \right) \ge H_{\min }^{\varepsilon '}\left( {\left. {{Z_A}} \right|E} \right) - lea{k_{EC}} - {\log _2}\frac{2}{{{\varepsilon _{cor}}}},
\end{equation}
where $H_{\min }^{\varepsilon '}\left( {\left. {{Z_A}} \right|E} \right)$ is the smooth min-entropy after error correction. Based on the uncertainty relation in~\cite{49}, the lower bound of $H_{\min }^{\varepsilon '}\left( {\left. {{Z_A}} \right|E} \right)$ is given by
\begin{equation}\label{A3}
\begin{array}{l}
H_{\min }^{\varepsilon '}\left( {\left. {{Z_A}} \right|E} \right) \ge {s_Z} - H_{\min }^{\varepsilon '}\left( {\left. {{{Z'}_A}} \right|{{Z'}_B}} \right)\\
{\kern 1pt} {\kern 1pt} {\kern 1pt} {\kern 1pt} {\kern 1pt} {\kern 1pt} {\kern 1pt} {\kern 1pt} {\kern 1pt} {\kern 1pt} {\kern 1pt} {\kern 1pt} {\kern 1pt} {\kern 1pt} {\kern 1pt} {\kern 1pt} {\kern 1pt} {\kern 1pt} {\kern 1pt} {\kern 1pt} {\kern 1pt} {\kern 1pt} {\kern 1pt} {\kern 1pt} {\kern 1pt} {\kern 1pt} {\kern 1pt} {\kern 1pt} {\kern 1pt} {\kern 1pt} {\kern 1pt} {\kern 1pt} {\kern 1pt} {\kern 1pt} {\kern 1pt} {\kern 1pt}{\kern 1pt}{\kern 1pt}{\kern 1pt}{\kern 1pt}{\kern 1pt}{\kern 1pt}{\kern 1pt}{\kern 1pt}{\kern 1pt} {\kern 1pt} {\kern 1pt} {\kern 1pt} {\kern 1pt} {\kern 1pt} {\kern 1pt} {\kern 1pt} {\kern 1pt} {\kern 1pt} {\kern 1pt} {\kern 1pt} {\kern 1pt} {\kern 1pt} {\kern 1pt}  \ge {s_Z}\left[ {1 - h\left( {\overline {E_{ph}^Z} } \right)} \right]
\end{array}
\end{equation}
Here, the upper bound of the phase error rate $\overline {E_{ph}^Z}$ is determined by Eq. (\ref{18}).
The secret key is ${\varepsilon _{\sec }}$-secret for a length $l$
\begin{equation}\label{A4}
l = \left[ {H_{\min }^{\varepsilon '}\left( {\left. {{Z_A}} \right|E'} \right) - 2{{\log }_2}\frac{1}{{2\nu }}} \right],
\end{equation}
where $\nu  = \frac{1}{2}\sqrt {{2^{l - H_{\min }^{\varepsilon '}\left( {\left. {{Z_A}} \right|E'} \right)}}}$. Hence, we obtain the secrecy
\begin{equation}\label{A5}
{\varepsilon _{\sec }} = 4\left( {\overline \varepsilon   + \underline \varepsilon  } \right) + 2\varepsilon ' + \varepsilon '' + \nu  + {\varepsilon _{PA}}.
\end{equation}
In order to get the secrecy in our protocol, the error terms are all fixed to a constant
\begin{equation}\label{A6}
\varepsilon ' = \varepsilon '' = \nu  = \overline \varepsilon   = \underline \varepsilon   = {\varepsilon _{PA}} = {\varepsilon _1},
\end{equation}
thus ${\varepsilon _{\sec }} = 13{\varepsilon _1}$.
\fontsize{10pt}{10pt}

\section{Statistical fluctuation models}
\fontsize{10pt}{10pt}
In the main text, we provide the finite-key analysis with the improved Chernoff bound. As a comparison, we utilize Hoeffding's inequality~\cite{37} and the multiplicative Chernoff bound~\cite{39} to acquire the tightest key rate bound. Here, we follow the definitions given in the main text. Let ${\tau _1},{\tau _2}, \ldots ,{\tau _n}$ be a set of $n$ independent Bernoulli random variables that ${\tau _i} \in \left\{ {0,1} \right\}$ and $\Pr \left( {{\tau _i} = 1} \right) = {P_i}$. Let $\tau  = \sum\nolimits_{i = 1}^n {{\tau _i}}$ denote an observed outcome. The mean value of the set is denoted as ${\tau ^ * } = \sum\nolimits_{i = 1}^n {{P_i}}$.

By applying Hoeffding's inequality, the upper bound and lower bound of the expectation value ${{\tau ^ * }}$ can be obtained by
\begin{equation}\label{B1}
\begin{split}
&\overline {{\tau ^ * }}  = \tau  + {{\hat \Delta }_H},\\
&\underline {{\tau ^ * }}  = \tau  - {\Delta _H},
\end{split}
\end{equation}
with failure probabilities ${{\hat \varepsilon }_H}$ and ${\varepsilon _H}$ separately, where ${{\hat \Delta }_H} = {g_H}\left( {N,{{\hat \varepsilon }_H}} \right)$, ${\Delta _H} = {g_H}\left( {N,{\varepsilon _H}} \right)$ and ${g_H}\left( {x,y} \right) = \sqrt {x/2\ln \left( {1/y} \right)}$.

In order to employ the multiplicative Chernoff bound, we are supposed to calculate the the lower bound of ${{\tau ^ * }}$ with the Hoeffding's inequality with a failure probability ${\varepsilon _H}$ at first. Then we denote ${\varepsilon _C}$ and ${{\hat \varepsilon }_C}$ as the failure probabilities of the upper bound and lower bound estimation of the Chernoff bound. Consider the following conditions: (1) ${\left( {2{{\hat \varepsilon }_C}^{ - 1}} \right)^{1/\underline {{\tau ^ * }} }} \le {e^{9/32}}$. (2) ${\varepsilon _C}^{ - 1/\underline {{\tau ^ * }} } < {e^{1/3}}$. If both conditions are satisfied, the upper bound and lower bound of the expectation value ${{\tau ^ * }}$ can be given by
\begin{equation}\label{B2}
\begin{split}
&\overline {{\tau ^ * }}  = \tau  + {{\hat \Delta }_C},\\
&\underline {{\tau ^ * }}  = \tau  - {\Delta _C},
\end{split}
\end{equation}
with failure probabilities ${{\hat \varepsilon }_C}$ and ${\varepsilon _C}$, where ${{\hat \Delta }_C} = {g_C}\left( {\tau ,{{\hat \varepsilon }_C}^4/16} \right)$, ${\Delta _C} = {g_C}\left( {\tau ,{\varepsilon _C}^{3/2}} \right)$ and ${g_C}\left( {x,y} \right) = \sqrt {2x\ln \left( {1/y} \right)}$. Because we have employed the Hoeffding's inequality to test the two conditions, the overall failure probability that the expectation value ${{\tau ^ * }}$ cannot be bounded by the multiplicative Chernoff bound is given by ${\varepsilon _H} + {\varepsilon _C} + {{\hat \varepsilon }_C}$. In this way, the expectation values estimated by variant simulation models are different. By comparing the performance of the final key rates, we can define the tightest security bound in our practical TF-QKD protocol.
\fontsize{10pt}{10pt}
\section{Estimation of ${Y_{nm}}$}
\fontsize{10pt}{10pt}
In this section, we provide the derivation of the upper bounds of ${Y_{0,0}}$, ${Y_{2,0}}$, ${Y_{0,2}}$ and ${Y_{1,1}}$ utilized in Eq. (\ref{16}) with the decoy-state method proposed in ~\cite{36}. In the main text, we have obtained the upper bounds and lower bounds of the expectation values $s_{AB}^{ * ,X}$ with the observables $s_{AB}^X$. Together with Eq. (\ref{9}), we have that
\begin{equation}\label{C1}
\frac{{{e^{{\mu _A} + {\mu _B}}}\underline {s_{AB}^{ * ,X}} }}{{{P_A}{P_B}{N_X}}} \le \sum\limits_{n,m} {\frac{{\mu _A^n\mu _B^m}}{{n!m!}}{Y_{nm}}}  \le \frac{{{e^{{\mu _A} + {\mu _B}}}\overline {s_{AB}^{ * ,X}} }}{{{P_A}{P_B}{N_X}}}.
\end{equation}
We define $T_{AB}^{ * ,X}: = {e^{{\mu _A} + {\mu _B}}}{\left( {{P_A}{P_B}{N_X}} \right)^{ - 1}}s_{AB}^{ * ,X}$. Then, Eq. (\ref{C1}) can be rewritten as
\begin{equation}\label{C2}
\underline {T_{AB}^{ * ,X}}  \le \sum\limits_{n,m} {\frac{{\mu _A^n\mu _B^m}}{{n!m!}}{Y_{nm}}}  \le \overline {T_{AB}^{ * ,X}},
\end{equation}
where
\begin{equation}\label{C3}
\begin{split}
\underline {T_{AB}^{ * ,X}}  = {e^{{\mu _A} + {\mu _B}}}{\left( {{P_A}{P_B}{N_X}} \right)^{ - 1}}\underline {s_{AB}^{ * ,X}},\\
\overline {T_{AB}^{ * ,X}}  = {e^{{\mu _A} + {\mu _B}}}{\left( {{P_A}{P_B}{N_X}} \right)^{ - 1}}\overline {s_{AB}^{ * ,X}}.
\end{split}
\end{equation}

Here, the decoy-state method proposed in ~\cite{37} can be directly applied to Eq. (\ref{C2}). The upper bound of ${Y_{1,1}}$ can be obtained by
\begin{equation}\label{C4}
\overline {{Y_{1,1}}}  = \frac{{\overline {{\Gamma _{1,1}}} }}{{{{\left( {{\mu _0} - {\mu _1}} \right)}^2}}},
\end{equation}
where
\begin{equation}\label{C5}
\overline {{\Gamma _{1,1}}}  = \overline {T_{00}^{ * ,X}}  + \overline {T_{11}^{ * ,X}}  - \underline {T_{01}^{ * ,X}}  - \underline {T_{10}^{ * ,X}}.
\end{equation}
\begin{widetext}
The upper bound of ${Y_{2,0}}$ and ${Y_{0,2}}$ are given by
\begin{equation}\label{C6}
\begin{split}
\overline {{Y_{2,0}}}  = \frac{{2\left[ {\left( {{e^{{\mu _0}}} - {e^{{\mu _1}}}} \right)\left( {{\mu _0} - {\mu _1} + {\mu _1}{e^{{\mu _0}}} - {\mu _0}{e^{{\mu _1}}}} \right) - \underline {{\Gamma _{2,0}}} } \right]}}{{\left( {{\mu _0} + {\mu _1}} \right){{\left( {{\mu _0} - {\mu _1}} \right)}^2}}},\\
\overline {{Y_{0,2}}}  = \frac{{2\left[ {\left( {{e^{{\mu _0}}} - {e^{{\mu _1}}}} \right)\left( {{\mu _0} - {\mu _1} + {\mu _1}{e^{{\mu _0}}} - {\mu _0}{e^{{\mu _1}}}} \right) - \underline {{\Gamma _{0,2}}} } \right]}}{{\left( {{\mu _0} + {\mu _1}} \right){{\left( {{\mu _0} - {\mu _1}} \right)}^2}}},
\end{split}
\end{equation}
where
\begin{equation}\label{C7}
\begin{split}
\underline {{\Gamma _{2,0}}}  = {\mu _1}\underline {T_{00}^{ * ,X}}  + {\mu _0}\underline {T_{11}^{ * ,X}}  - {\mu _0}\overline {T_{01}^{ * ,X}}  - {\mu _1}\overline {T_{10}^{ * ,X}},\\
\underline {{\Gamma _{0,2}}}  = {\mu _1}\underline {T_{00}^{ * ,X}}  + {\mu _0}\underline {T_{11}^{ * ,X}}  - {\mu _1}\overline {T_{01}^{ * ,X}}  - {\mu _0}\overline {T_{10}^{ * ,X}}.
\end{split}
\end{equation}

In order to obtain the upper bound of ${Y_{0,0}}$, we need the lower bound of ${Y_{2,2}}$ and the upper bounds of ${Y_{n,0}}$ and ${Y_{0,m}}$.

The upper bounds of ${Y_{n,0}}$ and ${Y_{0,m}}$ are given by
\begin{equation}\label{C8}
\begin{split}
&\overline {{Y_{n,0}}}  = \frac{{n!\left[ {\left( {{e^{{\mu _0}}} - {e^{{\mu _1}}}} \right)\left( {{\mu _0} - {\mu _1} + {\mu _1}{e^{{\mu _0}}} - {\mu _0}{e^{{\mu _1}}}} \right) - \underline {{\Gamma _{2,0}}} } \right]}}{{\left( {{\mu _0} - {\mu _1}} \right)\left( {\mu _0^n - \mu _1^n} \right)}},\\
&\overline {{Y_{0,m}}}  = \frac{{m!\left[ {\left( {{e^{{\mu _0}}} - {e^{{\mu _1}}}} \right)\left( {{\mu _0} - {\mu _1} + {\mu _1}{e^{{\mu _0}}} - {\mu _0}{e^{{\mu _1}}}} \right) - \underline {{\Gamma _{0,2}}} } \right]}}{{\left( {{\mu _0} - {\mu _1}} \right)\left( {\mu _0^m - \mu _1^m} \right)}}.
\end{split}
\end{equation}

The lower bound of ${Y_{2,2}}$ is given by
\begin{equation}\label{C9}
\underline {{Y_{2,2}}}  = \frac{{4\left[ {\overline {{\Gamma _{1,1}}}  - {{\left( {{e^{{\mu _0}}} - {e^{{\mu _1}}}} \right)}^2}} \right]}}{{{{\left( {\mu _0^2 - \mu _1^2} \right)}^2}}} + 1,
\end{equation}
where
\begin{equation}\label{C10}
\underline {{\Gamma _{1,1}}}  = \underline {T_{00}^{ * ,X}}  + \underline {T_{11}^{ * ,X}}  - \overline {T_{01}^{ * ,X}}  - \overline {T_{10}^{ * ,X}}.
\end{equation}

Then the upper bound of ${Y_{0,0}}$ can be obtained by
\begin{equation}\label{C11}
\begin{split}
\overline {{Y_{0,0}}}  = &\frac{{\overline {{\Gamma _{0,0}}} }}{{{{\left( {{\mu _0} - {\mu _1}} \right)}^2}}} + \frac{{{\mu _0}{\mu _1}\left[ {12\left( {{\mu _0} - {\mu _1}} \right)\left( {\overline {{Y_{2,0}}}  + \overline {{Y_{0,2}}} } \right) + 4\left( {\mu _0^2 - \mu _1^2} \right)\left( {\overline {{Y_{3,0}}}  + \overline {{Y_{0,3}}} } \right) + \left( {\mu _0^3 - \mu _1^3} \right)\left( {\overline {{Y_{4,0}}}  + \overline {{Y_{0,4}}} } \right)} \right]}}{{24\left( {{\mu _0} - {\mu _1}} \right)}}\\
 &+ \frac{{{\mu _1}\left( {24{e^{{\mu _0}}} - \mu _0^4 - 4\mu _0^3 - 12\mu _0^2 - 24} \right)}}{{12\left( {{\mu _0} - {\mu _1}} \right)}} - \frac{{{\mu _0}\left( {24{e^{{\mu _1}}} - \mu _1^4 - 4\mu _1^3 - 12\mu _1^2 - 24} \right)}}{{12\left( {{\mu _0} - {\mu _1}} \right)}} - \frac{{\mu _0^2\mu _1^2\underline {{Y_{2,2}}} }}{4},
\end{split}
\end{equation}
where
\begin{equation}\label{C12}
\overline {{\Gamma _{0,0}}}  = \mu _1^2\overline {T_{00}^{ * ,X}}  + \mu _0^2\overline {T_{11}^{ * ,X}}  - {\mu _0}{\mu _1}\left( {\underline {T_{01}^{ * ,X}}  + \underline {T_{10}^{ * ,X}} } \right).
\end{equation}
The upper bound and lower bound of $T_{AB}^{ * ,X}$ can be obtained by combining Eq. (\ref{C3}) with the observables $s_{AB}^X$.
\fontsize{10pt}{10pt}

\section{Linear channel loss model}
\fontsize{10pt}{10pt}
In order to perform our numerical simulation step, we adopt the linear channel loss model presented in ~\cite{36} with some modifications. The observed values which can be obtained in real experiments are simulated below. The overall loss between Alice (Bob) and Charlie is $\sqrt \eta$ accounting for both channel loss and detection efficiencies of the two detectors ($L$ and $R$). In the original practical TF-QKD protocol~\cite{36}, the gains and error rates are derived according to which detector ($L$ or $R$) clicks. Due to the symmetrical assumption of our system, we find that the difference between the total successful detection events of two detectors are negligible. Therefore, the total number of successful detection events when both parties choose $Z$-basis at the same time is denoted as
\begin{equation}\label{D1}
\begin{aligned}
{s_Z} = {N_Z}\left[ {\left( {1 - {P_d}} \right)\left( {{e^{ - \sqrt \eta  {\mu _Z}\cos \alpha \cos \beta }}
+ {e^{\sqrt \eta  {\mu _Z}\cos \alpha \cos \beta }}} \right){e^{ - \sqrt \eta  {\mu _Z}}} - 2{{\left( {1 - {P_d}} \right)}^2}{e^{ - 2\sqrt \eta  {\mu _Z}}}} \right],
\end{aligned}
\end{equation}
where $\alpha$ and $\beta$ are the polarization and phase misalignments of the signals from Alice and Bob after travelling through the optical channel,  ${N_Z}$ (${N_X}$) is the total number of pulses when both parties choose the $Z$-basis ($X$-basis).

The successful detection events in the $X$-basis $s_{AB}^X$ can be obtained by
\begin{equation}\label{D2}
\begin{aligned}
s_{AB}^X = 2{P_{{\mu _A}}}{P_{{\mu _B}}}{N_X}\left( {1 - {P_d}} \right)\left[ {\left( {{P_d} - 1} \right){e^{ - \sqrt \eta  \left( {{\mu _A} + {\mu _B}} \right)}} + {e^{ - \sqrt \eta  \left( {{\mu _A} + {\mu _B}} \right)/2}}{I_0}\left( {\sqrt {\eta {\mu _A}{\mu _B}} \cos \alpha } \right)} \right],
\end{aligned}
\end{equation}
where ${I_0}(x)$ is the modified Bessel function of the first kind.

The bit error rates in the $Z$-basis are given by
\begin{equation}\label{D3}
\begin{aligned}
{E_\mu ^Z} = {{{N_Z}\left( {1 - {P_d}} \right)\left[ {{e^{\sqrt \eta  {\mu _Z}\cos \alpha \cos \beta }} - \left( {1 - {P_d}} \right){e^{ - \sqrt \eta  {\mu _Z}\cos \alpha \cos \beta }}} \right]{e^{ - \sqrt \eta  {\mu _Z}}}} \mathord{\left/
 {\vphantom {{{N_Z}\left( {1 - {P_d}} \right)\left[ {{e^{\sqrt \eta  {\mu _Z}\cos \alpha \cos \beta }} - \left( {1 - {P_d}} \right){e^{ - \sqrt \eta  {\mu _Z}\cos \alpha \cos \beta }}} \right]{e^{ - \sqrt \eta  {\mu _Z}}}} {{n_Z}}}} \right.
 \kern-\nulldelimiterspace} {{s_Z}}},
 \end{aligned}
\end{equation}
\end{widetext}
\fontsize{10pt}{10pt}
\end{appendix}

\bibliographystyle{apsrev4-2}

\end{document}